\documentclass[preprint]{emulateapj} 

\usepackage{natbib}
\usepackage{hyperref}
\hypersetup{colorlinks,citecolor=Blue,linkcolor=Red,urlcolor=Blue}
\usepackage{amsmath}
\usepackage{empheq}
\usepackage[usenames,dvipsnames]{color}
\allowdisplaybreaks 

\newcommand{\icarus}{Icarus}

\begin{document}
 
\title{Spin-Spin Coupling in the Solar System}  
\author{Konstantin Batygin$^1$ \& Alessandro Morbidelli$^2$} 
%\author{K \& Morby} 

\affil{$^1$Division of Geological and Planetary Sciences, California Institute of Technology, Pasadena, CA 91125} 
\affil{$^2$Departement Lagrange, Observatoire de la C${\mathrm{\hat{o}}}$te d'Azur, 06304 Nice, France}
\email{kbatygin@gps.caltech.edu}

\begin{abstract} 

The richness of dynamical behavior exhibited by the rotational states of various solar system objects has driven significant advances in the theoretical understanding of their evolutionary histories. An important factor that determines whether a given object is prone to exhibiting non-trivial rotational evolution is the extent to which such an object can maintain a permanent aspheroidal shape, meaning that exotic behavior is far more common among the small body populations of the solar system. Gravitationally bound binary objects constitute a substantial fraction of asteroidal and TNO populations, comprising systems of triaxial satellites that orbit permanently deformed central bodies. In this work, we explore the rotational evolution of such systems with specific emphasis on quadrupole-quadrupole interactions, and show that for closely orbiting, highly deformed objects, both prograde and retrograde spin-spin resonances naturally arise. Subsequently, we derive capture probabilities for leading order commensurabilities and apply our results to the illustrative examples of (87) Sylvia and (216) Kleopatra asteroid systems. Cumulatively, our results suggest that spin-spin coupling may be consequential for highly elongated, tightly orbiting binary objects. 

\end{abstract} 

\maketitle

\section{Introduction}

The rotation of most natural satellites of the solar system is synchronous with their orbital periods \citep{Peale1999,MD99}. Beginning with the seminal study of \citet{1879Obs.....3...79D,1880Natur..21..235D}, this fact has long been understood to be a result of dissipative planet-satellite interactions \citep{1964Sci...145..881M,1964RvGSP...2..661K}. With considerable emphasis placed on describing the rotational evolution of the Moon \citep{1966RvGSP...4..411G}, the wide-spread applicability of spin-orbit synchronization as the end-state of tidal evolution was generally accepted as of half a century ago \citep{1965Sci...150.1717L,1965Natur.206R1240P}. However, the discovery of Mercury's a-synchronous spin (\citealt{1965Natur.206Q1240P}; see also \citealt{2007Sci...316..710M}) overturned the conventional understanding of tidal evolution of the time \citep{1988merc.book..461P}. 

Shortly after the observational revelation, it had become clear that Mercury's 59-day rotation period is a natural consequence of \textit{spin-orbit coupling}, an effect that qualitatively alters the isolated process of tidal de-spinning \citep{1965Natur.208..575C,1966AJ.....71..425G,1968ARA&A...6..287G,1966ApJ...145..296C}. In turn, the added insight has been instrumental to understanding the dynamical evolution of natural satellites \citep{1984Icar...58..137W,1987AJ.....94.1350W,2004AJ....128..484W,1987Natur.328..227M,1988Icar...76..295D,1995Icar..117..149B,2005Icar..176..224K,2009Icar..204..254T,2010Icar..207..732C} as well as the dramatic history of the Moon itself \citep{1979M&P....20..301M,1980M&P....23..185M,1986sate.book.....B,1994AJ....108.1943T,1998AJ....115.1653T,2004ARA&A..42..441C,2012Icar..219..241A,2012Sci...338.1047C,2013MNRAS.434L..21M}.

Spin-orbit coupling arises from gravitational torques exerted onto a permanently deformed satellite by the central object. Such modulation is particularly consequent when the torques accumulate coherently (i.e. when there exists a near-rational relationship between the orbital and rotational frequencies) and strongly depends on the degree of triaxiality of the satellite (see Ch. 5 of \citealt{MD99}). Therefore, exotic rotational states are naturally expected to be more prevalent among the solar system's small body populations, since typical object shapes become increasingly more irregular as mass is decreased\footnote{As an example, consider the fact that the asphericity of Mercury ($R = 2440$ km) is of order $(\mathcal{B}-\mathcal{A})/\mathcal{C} \sim 10^{-4}$, whereas that of Hyperion ($R = 135$ km) is $(\mathcal{B}-\mathcal{A})/\mathcal{C} \sim 0.25$ \citep{MD99,1987AJ.....94.1350W}.} (see for example \citealt{2010Icar..208..395T,2011Icar..212..649B} and the references therein).

Evidence for complex spin dynamics among sub-planetary objects exists within the current observational census. Specifically, Hyperion (a small satellite of Saturn) as well as Nix and Hydra (minor satellites of Pluto) are observationally inferred to rotate chaotically \citep{1989AJ.....97..570K,ShowHam}. Besides the highly unusual satellite figures, in the case of Hyperion, chaos is facilitated by an eccentric orbit \citep{1984Icar...58..137W,1987AJ.....94.1350W}, whereas in the Plutonian system, irregular motion stems from perturbations due to another satellite, namely Charon \citep{Correia2015,1967AJ.....72..662G}. While the families of spin-orbit interactions at play in the aforementioned examples are distinct, these systems share a common feature: the central objects in both cases are almost perfectly spherical. 

Not all satellites in the solar system orbit spherical bodies. In particular, the nearly three-decade old inquiry of \citealt{1989aste.conf..643W} - ``Do asteroids have satellites?" has now been definitively and positively answered (see \citealt{2002aste.conf..289M} - ``Asteroids do have satellites"). In fact, asteroidal and trans-Neptunian binaries comprise a non-negligible fraction of the overall small body population. Among near-Earth asteroids, $\sim 15 \%$ are thought to be binaries, with a similar fraction corresponding to small ($R \lesssim 10$ km) main-belt asteroids \citep{1996Natur.381...51B,2006Icar..181...63P}. Among larger asteroids ($10$ km $\lesssim R \lesssim 100$ km), the binary fraction is probably somewhat smaller (e.g. $\sim$ few \%), similar to that of the dynamically hot component of the Kuiper belt \citep{2002aste.conf..289M,2006AREPS..34...47R,2008ssbn.book..345N}. The binary fraction among the cold classical population of the Kuiper belt is considerably more enhanced and is thought to be of order $\sim 25\%$ \citep{2006AJ....131.1142S}.

There are a few different formation channels that may be associated with binary populations occupying different parts of the solar system. A process most relevant to the formation of near-Earth and small main-belt asteroidal binaries appears to be rotational fission, facilitated by radiative torques i.e. the YORP effect \citep{2000Icar..145..301B,2002aste.conf..395B,2007Icar..189..370S,2008Natur.454..188W,2010Natur.466.1085P}. The primary mode of formation of larger binary asteroids is likely associated with (sub-)catastrophic impacts \citep{1996Icar..120..212D,2004Icar..170..243D,1997P&SS...45..757D,2001Sci...294.1696M,2002aste.conf..527S}. In trans-Neptunian space, dynamical capture \citep{2002Natur.420..643G,2005MNRAS.360..401A,2008ssbn.book..345N}, collisions \citep{2005Icar..173..342P,2005Sci...307..546C,2006Natur.439..946S,2009ApJ...700.1242S,2010ApJ...714.1789L}, and fission during gravitational collapse \citep{2010AJ....140..785N} are believed to dominate binary formation. 

Irrespective of the exact formation mechanism, highly deformed objects smaller than $R \lesssim 100$ km can be found throughout the solar system. Thus, a considerable fraction of all minor object binaries may comprise triaxial satellites that orbit permanently deformed primaries \citep{2006Icar..185...39M}. Real life examples of such systems include the triple asteroid systems (87) Sylvia and (216) Kleopatra \citep{2000Sci...288..836O,2001IAUC.7588....1B,2005Natur.436..822M,2011Icar..211.1022D}. These systems are comprised of large ($R_p \sim 200$ km) primary objects, orbited by minuscule ($R_s \sim$ few km) moons. The orbital separation of the moons is of order $a \sim$ few $R_p$ and the longest dimension of the primaries is roughly twice that of the shortest dimension. The tidal spin down timescale associated with these satellites is of order tens of thousands of years \citep{GoldreichSari2009}. 

How is the rotational evolution of the small moons affected by the highly irregular shapes of the central objects? Qualitatively speaking, the gravitational potential of a strongly triaxial body harbors a substantial quadrupolar component. As a result, during the course of tidal spin-down or spin-up, the rotation rate of the moon may become commensurate with the evolution frequency of the quadrupolar part of the potential, which in turn varies with the spin of the primary. In other words, rotational evolution of multiple small-body systems can be subject to \textit{spin-spin coupling}. 

In this work, we will explore the spin-spin coupling effect quantitatively. We note that observational characterization of multiple asteroid systems is a relatively young field (the first triple asteroid, (87) Sylvia was discovered a decade ago - see \citealt{2005Natur.436..822M}), and direct measurements of the rotation rates of minor moons do not yet exist. Accordingly, the primary goal of this paper is to provide broad predictions regarding yet uncharacterized modes of rotational evolution, rather than to perform a detailed study. Correspondingly, simplicity is favored over realism throughout the performed calculations. 

The paper is organized as follows. In section 2, we analytically derive the governing equations for low-order spin-spin resonances, and consider their stability. In section 3, we compute the associated probabilities of capture. In section 4, we present illustrative applications of the obtained results to the multiple asteroid systems mentioned above. We conclude and discuss our findings in section 5.

\section{Resonant Torques}

\begin{figure}
\includegraphics[width=1\columnwidth]{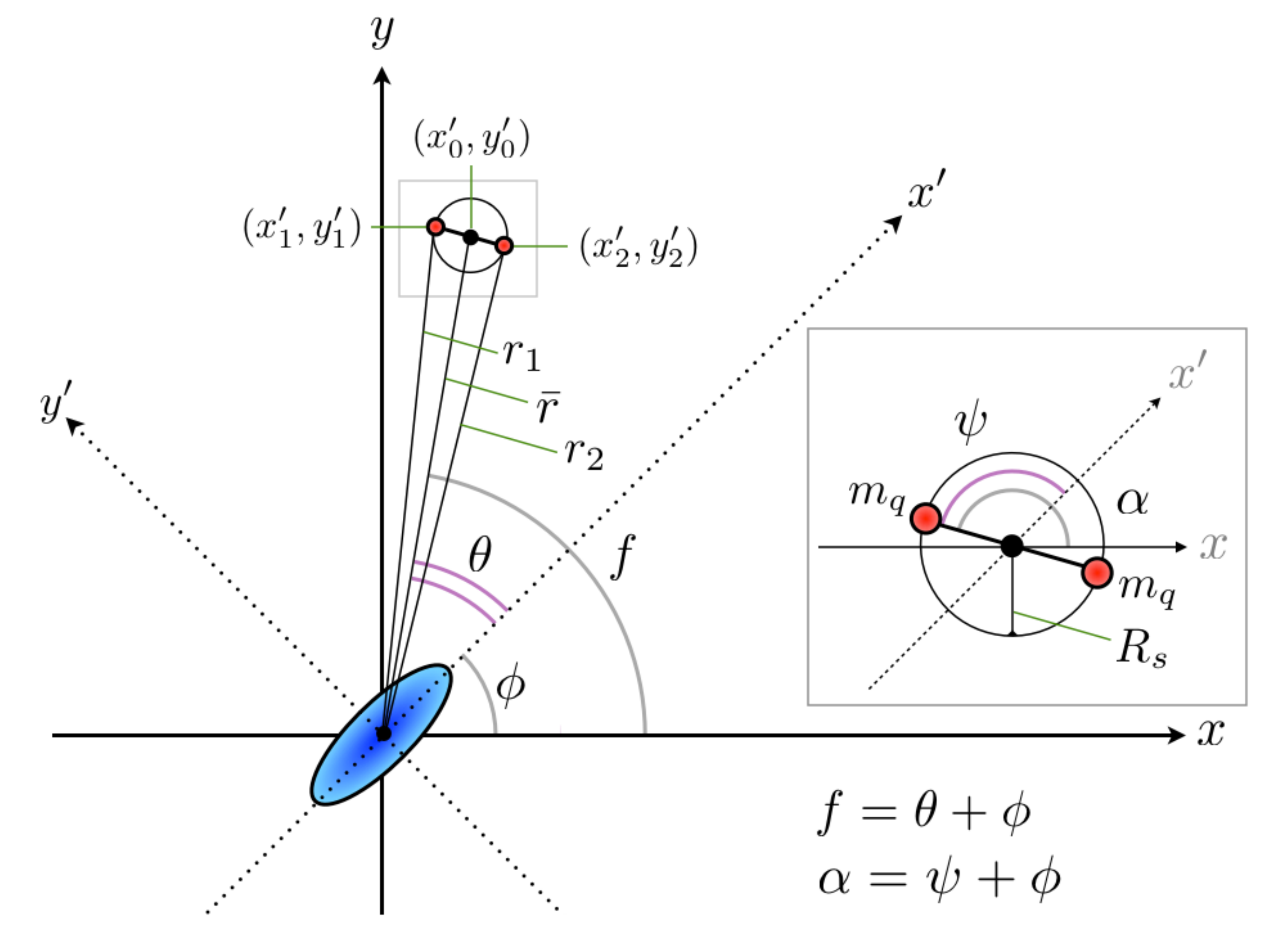}
\caption{The geometrical setup associated with the presented calculations. $\alpha$ refers to the physical orientation of the satellite, while $\phi$ corresponds to the rotation of the primary. $f$ denotes the true longitude of the satellite. In a non-inertial (primed) coordinate system correlating with the primary, $\alpha$ and $f$ are replaced by the angles $\psi$ and $\theta$ respectively.}
\label{setup}
\end{figure}

The coupling function we aim to derive here is one describing the rigid body dynamics of an aspherical secondary (satellite) of negligible mass, orbiting a rotating aspherical primary (planet/asteroid/KBO). For definitiveness, only the secondary's spin is assumed to evolve under the influence of the primary's gravitational potential. Furthermore, we assume that the spin vectors of the objects and the corresponding principal axes are normal to the orbital plane. In other words, the formulation of the problem is restricted and planar. Moreover, motivated by the nearly circular and planar orbits of (87) Sylvia and (216) Kleopatra systems \citep{2011Icar..211.1022D,2012AJ....144...70F}, we shall assume null eccentricities. 

\subsection{Geometrical Setup}

The geometrical setup of the problem is depicted in Figure (\ref{setup}). For a considerable fraction of the following derivation, we shall work in a rotating reference frame ($x',y'$), centered on the barycenter of the primary and synchronized with its spin, $\dot{\phi}$, such that the $x'$-axis is always aligned with the long axis of the primary ellipsoid. The radial line joining the secondary's barycenter position ($x_0',y_0'$) and the origin has the length $\bar{r}$ and makes an angle $\theta$ with the $x'$-axis. The angular position of the secondary in the rotating frame ($x',y'$) can be translated to that in an inertial frame ($x,y$) by adding the primary orientation angle to $\theta$. As such, their sum is simply interpreted as the secondary's true longitude $f = \theta + \phi$.

Following \citet{Hut1981}, we treat the secondary as a sphere of radius $R_s$, and represent its quadrupole moment with two diametrically opposed point masses, $m_q$, placed a distance $R_s$ away from the barycenter. In this formulation, the difference in the principal moments of inertia is interpreted as: $\mathcal{B}_s-\mathcal{A}_s = 2 m_q R_s^2$ \citep{MD99}, keeping in mind that $\mathcal{A}_s$ and $\mathcal{C}_s$ correspond to the long axis and the axis or rotation of the secondary, respectively. 

The positions of the masses $m_q$ in the rotating frame are arbitrarily taken to be ($x_1',y_1'$) and ($x_2',y_2'$), defining the distance between them and the primary's barycenter as $r_1$ and $r_2$ respectively. The angle made between the $x'$-axis and the line connecting ($x_1',y_1'$) to ($x_0',y_0'$) is labeled as $\psi$ while $\alpha = \psi + \phi$ is taken to be its inertial counterpart.  

\subsection{Derivation of the Torques}

With all relevant angles defined, let us now calculate the torque exerted on the secondary by the primary. We begin by expressing
\begin{align}
\label{x0y0}
(x_0',y_0') = (\bar{r} \cos{\theta}, \bar{r} \sin{\theta}),
\end{align}
and
\begin{align}
\label{x1y1}
(x_1',y_1') = (\bar{r} (\epsilon \cos{\psi} + \cos{\theta} ),\bar{r} (\epsilon \sin{\psi} + \sin{\theta} )),
\end{align}
where $\epsilon = R_s/\bar{r} \ll 1$. Additionally, we have
\begin{align}
\label{r1}
r_1 = \sqrt{\bar{r}^2+R_s^2+2\bar{r}R_s \cos{(\psi-\theta)}}.
\end{align}
Correspondent expressions for ($x_2',y_2'$) and $r_2$ can be obtained by replacing $\psi$ by $\psi - \pi$ in the above expressions. 

The torque on the secondary due to $m_q$ at ($x_1',y_1'$) is given by
\begin{align}
\label{torque1}
\mathbb{T}_1 = -m_q (x_1'-x_0',y_1'-y_0',0) \times \vec{\nabla} \mathcal{U} \big{|}_{(x_1',y_1')},
\end{align}
where $\mathcal{U}$ is the gravitational potential of the primary. To second order in the quantity $(R_p/\bar{r})$, the primary's potential can be approximated by MacCullagh's formula \citep{MD99}:
\begin{align}
\label{MacCullagh}
\mathcal{U} = &-\frac{G m_p}{r} - \frac{G (\mathcal{B}_p + \mathcal{C}_p - 2 \mathcal{A}_p ) x^2}{r^5} \nonumber \\
&+ \frac{G (\mathcal{C}_p + \mathcal{A}_p - 2 \mathcal{B}_p ) y^2 }{r^5},
\end{align}
where $\mathcal{A}_p, \mathcal{B}_p,$ and $\mathcal{C}_p$ are the primary's principal moments of inertia\footnote{Higher order expressions have been worked out by \citet{2007CeMDA..99..149A} and the references therein.}. Substitution of equations (\ref{x0y0}), (\ref{x1y1}), (\ref{r1}) and (\ref{MacCullagh}) into equation (\ref{torque1}) yields the following expression for the orbit-normal component of the torque (the other components being zero by construction):
\begin{align}
\label{torque1full}
\mathbb{T}_1&= -\frac{G m_q \epsilon }{8 \bar{r}^3 \left(\epsilon ^2+1+2 \epsilon  \cos (\theta -\psi ) \right)^{7/2}} \times \nonumber \\
\big( &- 42 \epsilon  (\mathcal{A}_p-\mathcal{B}_p) \sin (2 \theta ) \nonumber \\
 &-6 \epsilon  \left(2 \epsilon ^2-1\right) (\mathcal{A}_p-\mathcal{B}_p) \sin (2 \psi ) \nonumber \\
&- 6 \left(\epsilon ^2+1\right) (\mathcal{A}_p+\mathcal{B}_p-2 \mathcal{C}_p) \sin (\theta -\psi ) \nonumber \\
&+ 8 m_p \bar{r}^2 \left(\epsilon ^4+3 \epsilon ^2+1\right) \sin (\theta -\psi ) \nonumber \\
&-6 \epsilon  (\mathcal{A}_p+\mathcal{B}_p-2 \mathcal{C}_p) \sin (2 (\theta -\psi )) \nonumber \\ 
&+16 m_p \bar{r}^2 \epsilon  \left(\epsilon ^2+1\right) \sin (2 (\theta -\psi )) \nonumber \\
&+ 8 m_p \bar{r}^2 \epsilon ^2 \sin (3 (\theta -\psi )) \nonumber \\
&-15 (\mathcal{A}_p-\mathcal{B}_p) \sin (3 \theta -\psi ) \nonumber \\
&- 3 \epsilon ^2 (\mathcal{A}_p-\mathcal{B}_p) \sin (\theta -3 \psi ) \nonumber \\
&- 3 \left(13 \epsilon ^2-1\right) (\mathcal{A}_p-\mathcal{B}_p) \sin (\theta +\psi )\big).
\end{align}
As already mentioned above, the equation for $\mathbb{T}_2$ can simply be obtained by replacing $\psi$ in the above expression by $\psi - \pi$. 

While complete, the above expression for the torque is rather cumbersome. Consequently, rather than working with the full expression for the total quadrupole torque $\mathbb{T}_q = \mathbb{T}_1+ \mathbb{T}_2$, it is sensible to expand the torque as a power series in the small parameter $\epsilon$ and Fourier decompose the results. This procedure allows for a more straightforward study of the individual resonant harmonics. We take this approach below, limiting ourselves only to zeroth and first order resonances.

\subsection{Zeroth Order Resonances}
Setting $m_q \epsilon^2 = (\mathcal{B}_s-\mathcal{A}_s)/2 \bar{r}^2$ in the above expression, expanding to leading order in $\epsilon$ and switching to angles in the inertial frame, we obtain:
\begin{align}
\label{T0}
&\mathbb{T}_q^{(0)} = \frac{3 G m_p (\mathcal{B}_s-\mathcal{A}_s) \sin (2 (f-\alpha ))}{2 \bar{r}^3} \nonumber \\
&\times \left( 1 - \frac{5 (\mathcal{A}_p + \mathcal{B}_p - 2 \mathcal{C}_p)}{4 m_p \bar{r}^2} \right) \nonumber \\
&- \frac{21 G (\mathcal{B}_p-\mathcal{A}_p) (\mathcal{B}_s-\mathcal{A}_s) \sin (2 (\alpha -\phi ))}{16 \bar{r}^5} \nonumber \\
&- \frac{105 G (\mathcal{B}_p-\mathcal{A}_p) (\mathcal{A}_s-\mathcal{B}_s) \sin (4 f-2 (\alpha +\phi ))}{16 \bar{r}^5}.
\end{align}
The first harmonic in the above expression is responsible for conventional \textit{spin-orbit coupling} \citep{1966AJ.....71..425G}. In fact, if we take the primary to be completely spherical ($\mathcal{A}_p = \mathcal{B}_p = \mathcal{C}_p$), the above equation reduces to the well-known expression for torque exerted onto an aspherical satellite by a point-mass \citep{1988fcm..book.....D}. 

The other two harmonics in the expansion are responsible for \textit{spin-spin coupling}. In particular, both terms give rise to 1:1 spin-spin resonances. Note, however, that while the second term librates when the rotation rate of the secondary is commensurate with that of the primary (i.e. $\dot{\alpha} \simeq \dot{\phi}$), the third harmonic requires the rotation rates to have opposite signs. In a regime where the orbital frequency is well separated from the rotational frequencies (i.e. $\dot{f} \ll \dot{\phi}$), this harmonic resonates when the rotation rates of the primary and secondary are equal and opposite (i.e $\dot{\alpha} \simeq - \dot{\phi}$). In other words, in the limit where $f$ can be considered to be a slowly varying angle, the second term in equation (\ref{T0}) governs the \textit{prograde} 1:1 spin-spin resonance while the third term governs the \textit{retrograde} 1:1 spin-spin resonance. 

Incidentally, here a direct analogy between spin-spin resonances and mean-motion orbit-orbit resonances can be drawn. The critical angles of mean motion resonances also comprise combinations of differences between the quickly-varying mean longitudes and the slowly varying secular angles, namely the longitude of perihelion and the ascending node (see for example Ch.8 of \citealt{MD99}). Mean motion resonances are generally labeled according to the coefficients in front of the mean longitudes in the critical arguments. Consequently, in this work, we follow the established convention in labeling the spin-spin resonances.

\subsection{First Order Resonances}
Following the same procedure as for zeroth order torques, the Fourier decomposition of the first order torques take the form
\begin{align}
\label{T1}
\mathbb{T}_q^{(1)} &= \frac{21 \, \epsilon \, G (\mathcal{B}_p-\mathcal{A}_p) (\mathcal{B}_s-\mathcal{A}_s) \sin (\alpha -2 \phi + f)}{8 \bar{r}^5} \nonumber \\
&+ \frac{21 \, \epsilon \, G (\mathcal{B}_p-\mathcal{A}_p) (\mathcal{B}_s-\mathcal{A}_s) \sin (3 \alpha - 2 \phi - f)}{8 \bar{r}^5}.
\end{align}
Both of these terms govern spin-spin resonances. In particular the first term corresponds to the 1:2 resonance while the second corresponds to a 3:2 resonance. Note that all resonant terms obey D'Almbert rules: writing any critical argument as ($j_1 \alpha + j_2 \phi + j_3 f$), the sum of the coefficients $j$ is always zero. This is because the setup of the problem is invariant under rotation of the inertial frame. 

We can in principle continue the expansion and derive additional higher order torques. However, the strength of the resonances rapidly decreases. As a result, here we shall stop the expansion at first order, leaving further elaboration for future studies. 

\subsection{Stability of Spin-Spin Resonances}

As already mentioned above, for a spherical primary and a (nearly) circular orbit, the only stable end-state of tidal evolution is a 1:1 spin-orbit resonance, such as that observed for the Moon and numerous other planetary satellites. The presented calculation shows however, that this is no longer the case if both, the primary and the secondary are significantly non-spherical. Still, for an object to become permanently trapped in a spin-spin resonance, such a resonance must be stable (note that this is a separate issue from capture probabilities, which we consider below). As \citet{1966AJ.....71..425G} point out, the stability of a resonance is assured simply by requiring the restoring quadrupole torque to exceed the tidal torque i.e. $\mathbb{T}_q > \mathbb{T}_t$ when the critical argument of the resonance approaches $\pi$ or $0$ (depending on whether the resonance in question is prograde or retrograde). 

In the framework of the constant time-lag (CTL) tidal model \citep{1979M&P....20..301M,1980M&P....23..185M,Hut1981}, the relevant tidal torque reads
\begin{align}
\mathbb{T}_t = - 3k_{2_s} \tau \frac{G m_p^2 R_s^5}{\bar{r}^6} (\dot{\alpha} - n)
\label{CTLtorque}
\end{align}
where $k_{2_s} = 3/2(1+\bar{\mu})^{-1} \sim 10^{-3} (R_s/1000\rm{km})^2$ is the secondary's Love number ($\bar{\mu}$ is the dimensionless mean rigidity), $n = \dot{f}$ is the orbital frequency, and $\tau$ is the (small) time by which the tidal potential is assumed to lag the perturbing motion \citep{Peale1999}. Within the CTL model, the oft-quoted dissipation quality factor is related to the time lag by $Q_s^{-1} = 2 \tau |\dot{\alpha} - n|$ \citep{EfroimskyLainey2007}. 

We note that although the CTL tidal model has been utilized extensively in various astrophysical contexts (e.g. \citealt{Hut1981, 2009Icar..201....1C} and the references therien), it only serves as an adequate approximation to more realistic rheological models within a narrow parameter range. As an example, for a body obeying Maxwell rheology, the CTL description can be employed provided that the viscosity is smaller than \citep{Efroimsky2015}
\begin{align}
\nu < \frac{4 \pi}{57} \frac{G \rho_s^2 R_s^2}{|\dot{\alpha} - n|}.
\end{align}
Strictly speaking, this implies that the following results apply to objects with viscosities orders of magnitude below that of ice. However, in the spirit of simplicity, here we shall follow the (formally incorrect) convention of celestial mechanics and employ equation (\ref{CTLtorque}), leaving more complicated analyses for future work.

\begin{figure}[t]
\includegraphics[width=1\columnwidth]{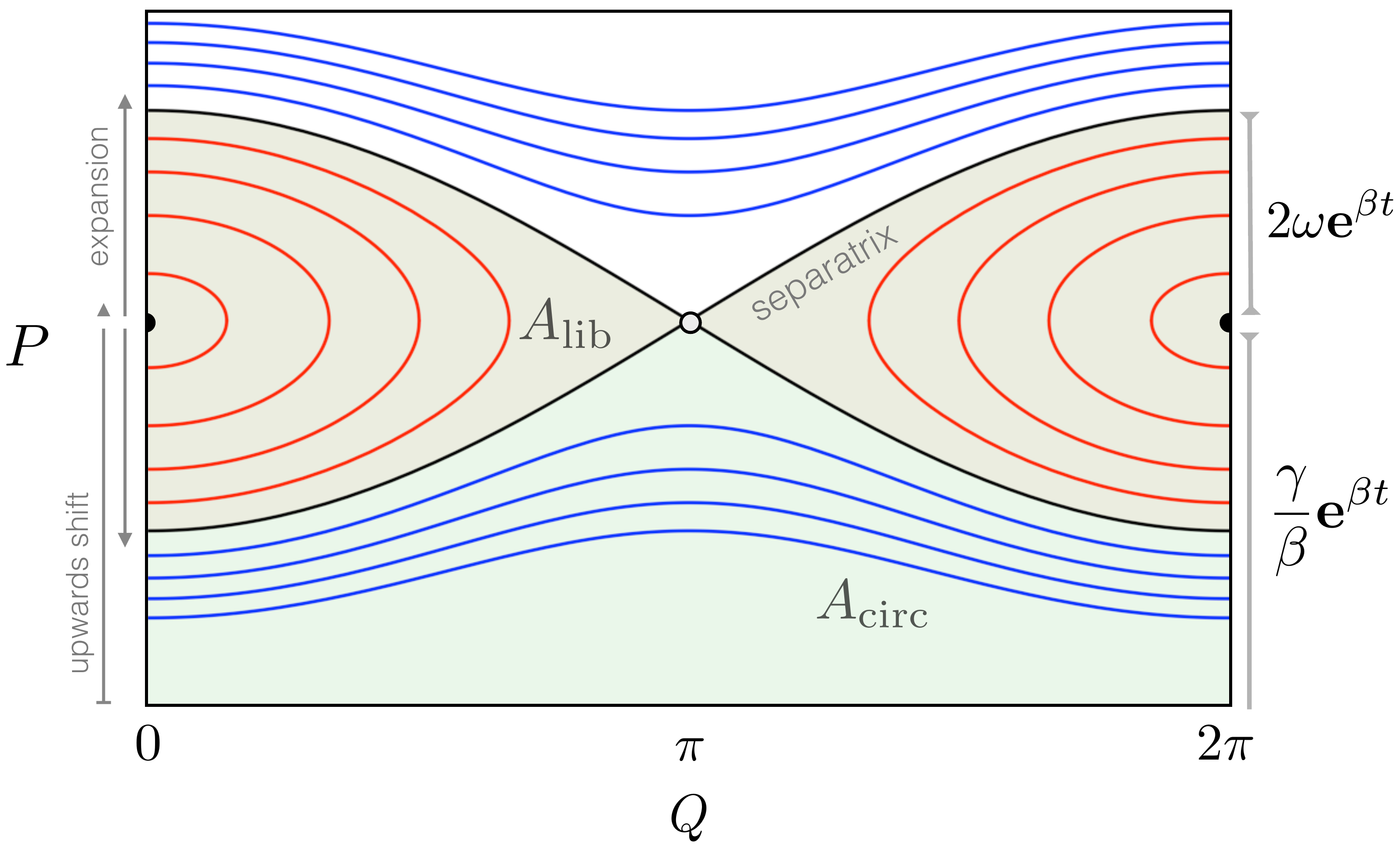}
\caption{Level curves of the Hamiltonian (\ref{K'}). The red curves denote librating trajectories while the blue curves denote circulating trajectories. The separatrix is shown as the black curve. The width of the resonance as well as the location of the fixed points are labeled. Note that both expand in time as $\propto \mathbf{e}^{\beta t}$. The stable fixed points are labeled with filled dots, while the unstable fixed point is shown with an empty dot. Librational and circulational phase-space areas are also labeled.}
\label{pendulum}
\end{figure}

Bearing in mind the form that quadrupole torques take from equations (\ref{T0}) and (\ref{T1}), the stability of spin-spin resonances is ensured when 
\begin{align}
&\left| \frac{3 k_{2_s} \tau m_p^2 R_s^5 (\dot{\alpha} - n)}{\xi \epsilon^p \bar{r} (\mathcal{B}_p-\mathcal{A}_p) (\mathcal{B}_s-\mathcal{A}_s)} \right| \nonumber \\
&= \left| \frac{3}{2 \xi} \frac{k_{2_s}}{Q_s} \frac{m_p}{m_s} \left( \frac{R_s}{R_p} \right)^2 \frac{\epsilon^{1-p}}{\lambda_s \lambda_p} \right|  \ll1,
\end{align}
where $\xi$ is the resonant torque coefficient (generally of order $\sim \rm{few} $ - e.g. 21/8 in the case of first order resonances), $p$ is the order of resonance and $\lambda = (\mathcal{B}-\mathcal{A})/mR^2$.

As an example, let us consider a $R_s \sim 100$km secondary orbiting a $R_s \sim 1000$km primary, naively setting $Q \sim 100$ \citep{1966Icar....5..375G}. For such a binary, first order resonances require $(\lambda_s \lambda_p) \sim 10^{-5}$ for stability. For a physically smaller binary ($R_s \sim 10$km secondary orbiting a $R_p \sim 100$km primary), the required value changes to $(\lambda_s \lambda_p) \sim 10^{-7}$, thanks to the Love number's dependence on the secondary's radius. The required $(\lambda_s \lambda_p)$ further decreases by a factor of $\epsilon$ for zeroth order resonances. Note, however, that if the secondary in question is a rubble-pile, tidal evolution proceeds at an enhanced rate as the mean rigidity gets replaced with $\bar{\mu}_{\rm{eff}}\rightarrow \sqrt{\bar{\mu}/Y}$ where $Y \sim 10^{-2}$ is the yield strain \citep{GoldreichSari2009}. Consequently, the critical $(\lambda_s \lambda_p)$ must also be somewhat higher for rubble piles. 

Collectively, the above arguments suggest that only small deviations from absolute sphericity of both bodies are required for low-order spin-spin resonances to stabilize the secondary's rotation against tidal de-spinning. As already mentioned above however, this fact alone does not guarantee that an encounter with a spin-spin resonance will result in capture. As a result, we shall consider spin-spin resonance capture probabilities in the next section.

\section{Capture Probabilities}

In the vicinity of a given spin-spin resonance, characterized by a critical angle ($j_1 \alpha + j_2 \phi + j_3 f$), all other quadrupole torques will average to zero over many circulation cycles. Consequently, it is sensible to consider averaged equations of motion where only a single resonant torque is retained:
\begin{align}
\label{eom}
\mathcal{C}_s \ddot{\alpha} &+ \frac{\xi \epsilon^p G (\mathcal{B}_p-\mathcal{A}_p) (\mathcal{B}_s-\mathcal{A}_s) \sin (j_1 \alpha + j_2 \phi + j_3 f)}{ \bar{r}^5} \nonumber \\
&+ 3k_{2_s} \tau \frac{G m_p^2 R_s^5}{\bar{r}^6} (\dot{\alpha} - n) = 0
\end{align}
Changing variables to $\upsilon =  (j_1 \alpha + j_2 \phi + j_3 f)$, the averaged equation of motion simplifies to that of a damped pendulum:
\begin{align}
\label{eompendulum}
\ddot{\upsilon}+\omega^2 \sin(\upsilon) + \beta \dot{\upsilon} + \gamma = 0.
\end{align}
In the above expression, a circular orbit has been implicitly assumed yielding the following constants:
\begin{align}
\omega^2 &= j_1 \xi \epsilon^p n^2 \lambda_p (R_p/\bar{r})^2 (\mathcal{B}_s-\mathcal{A}_s)/ \mathcal{C}_s \nonumber \\
\beta &= 3k_{2_s} \tau n^2 \epsilon^3 (m_p/m_s) (m_s R_s^2/\mathcal{C}_s ) \nonumber \\
\gamma &= j_2 \beta (n - \dot{\phi}).
\end{align}
Immediately, the argument made above about the stability of resonances can be understood as a criterion for existence of librating trajectories (and by extension, the existence of the separatrix). In other words, if $|\gamma| > \omega^2$, $\dot{\upsilon}$ can not reverse sign as a result of quadrupole torques, implying continuous circulation. Obviously, without a librational island in phase-space, capture into resonance cannot take place (recall however, that this is a weak criterion). Note further, that $\omega$ is the natural libration frequency of the resonant angle, $\upsilon$. 

Following \citet{HenrardBible}, the equation of motion (\ref{eompendulum}) can be understood as arising from a pendulum-like Hamiltonian
\begin{align}
\label{conspend}
\mathcal{H}(\upsilon, I) = \frac{1}{2} \left( I - \frac{\gamma}{\beta} \right)^2 - \omega^2 \cos(\upsilon),
\end{align}
where $I = \dot{\upsilon} - \gamma/\beta$ is the conjugated momentum, perturbed by dissipative forces:
\begin{align}
\label{eomdiss}
\dot{\upsilon} = \frac{\partial \mathcal{H}}{\partial{I}} \ \ \ \ \ \dot{I} = - \frac{\partial \mathcal{H}}{\partial{\upsilon}} - \beta I.
\end{align}
Note that unlike mean-motion resonances of celestial mechanics, Hamiltonian (\ref{conspend}) does not posses the d'Alembert characteristic (dependence of $\omega^2$ on the action; \citealt{SFMR}).

There exists a general procedure for incorporation of dissipative effects into the Hamiltonian framework which we shall utilize here (see \citealt{HenrardBible,2013PhRvL.110q4301G,Tsang} and the references therein).  Let the auxiliary system of equations
\begin{align}
\dot{\tilde{\upsilon}} = F_{1} \ \ \ \ \ \dot{\tilde{I}} = F_{2},
\end{align}
where $F_1$ and $F_2$ are dissipative forces and ($Q,P$) are initial conditions, be satisfied by the solutions $\tilde{\upsilon}(Q,P, t)$ and $\tilde{I}(Q, P, t)$. These solutions can be interpreted as a transformation from the original variables, ($\upsilon$,$I$) to ($Q$,$P$) via $\upsilon = \tilde{\upsilon}$ and $I = \tilde{I}$. In other words, the Hamiltonian flow can be envisioned as an initial condition to the dissipative solution. The new Hamiltonian, $\mathcal{K}$ that describes the system is then related to the old Hamiltonian, $\mathcal{H}$ by 
\begin{align}
\mathcal{K}(Q,P,t) = \mathcal{H}(\tilde{\upsilon}(Q,P,t),\tilde{I}(Q,P,t)),
\end{align}
while the new canonical time, $\sigma$ is defined by the Wronskian
\begin{align}
\label{wronskian}
\frac{dt}{d\sigma} = \left(\frac{\partial \tilde{\upsilon}}{\partial Q} \frac{\partial \tilde{I}}{\partial {P}} - \frac{\partial \tilde{\upsilon}}{\partial P} \frac{\partial \tilde{I}}{\partial {Q}}    \right) = \{\tilde{\upsilon},\tilde{I}\}_{(Q,P)},
\end{align}
where $\{\}$ signifies the Poisson bracket. 

\begin{figure}[t]
\includegraphics[width=1\columnwidth]{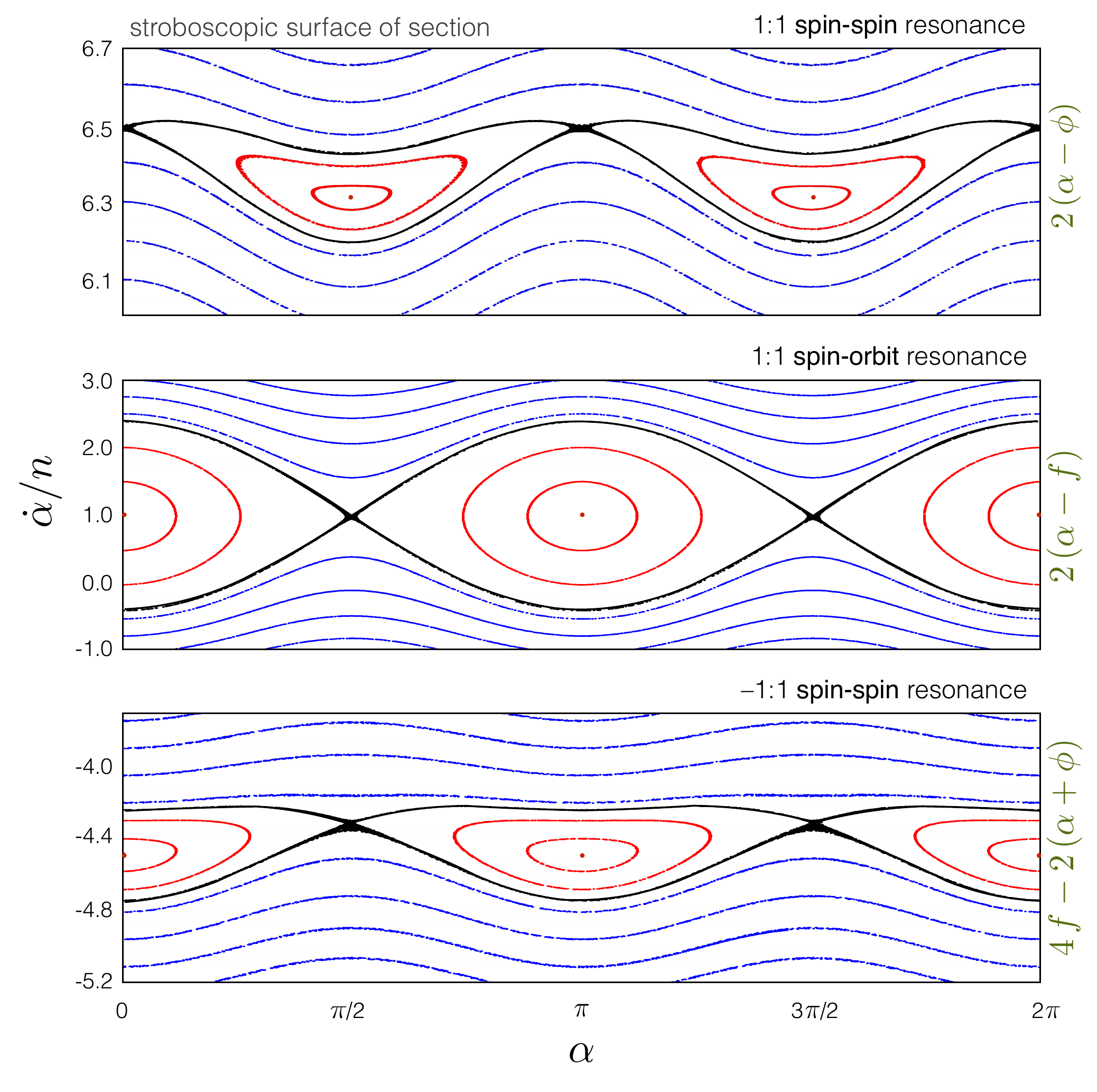}
\caption{A stroboscopic surface of section, showing the phase-space portrait of the inertial angle $\alpha$, corresponding to parameters characteristic of Remus, the inner satellite of the (87) Sylvia system. The parameters are chosen in accord to the observed properties of the system: $R_p/a = 0.2$, $\epsilon = 0.01$, $(\mathcal{B}_p-\mathcal{A}_p)/ (m_p R_p^2) = 0.19$. For the purposes of this figure, the same degree of triaxiality as that inferred for Nix is adopted i.e. $(\mathcal{B}_s-\mathcal{A}_s)/\mathcal{C}_s = 0.63$. The three panels of the figure show resonances associated with the three harmonics that appear in equation (\ref{T0}).}
\label{section}
\end{figure}

In the context of the spin-spin resonance problem, equation (\ref{eomdiss}) implies that $F_1 = 0$ while $F_2 = -\beta I$. Consequently, we are presented with the following relationships:
\begin{align}
\tilde{\upsilon} = Q \ \ \ \ \ \tilde{I} = P \mathbf{e}^{-\beta t}.
\end{align}
Accordingly, the new Hamiltonian takes the form
\begin{align}
\label{K}
\mathcal{K}(Q,P,t) = \frac{1}{2} \left(P \mathbf{e}^{-\beta t} - \frac{\gamma}{\beta}\right)^2 - \omega^2 \cos(Q)
\end{align}
The Poisson bracket (\ref{wronskian}) evaluates to $\{\tilde{\upsilon},\tilde{I}\}_{(Q,P)} = \mathbf{e}^{-\beta t}$, defining $\sigma =\mathbf{e}^{\beta t}/\beta$.

In the current form, the Hamiltonian (\ref{K}) yields equations of motion where the momentum $P$ is multiplied by an explicitly time-dependent function. Such formulation is awkward, and it is preferable to transfer the time-dependence to the coefficients. We can rescale the Hamiltonian (\ref{K}) by a time-dependent function $\chi(t)$ such that
\begin{align}
\mathcal{K'} = \chi(t) \mathcal{K} \ \ \ \ \ \ d \sigma' = \frac{d\sigma}{\chi(t)}.
\end{align}
Choosing $\chi(t) = \mathbf{e}^{2 \beta t}$, we obtain: 
\begin{align}
\label{K'}
\mathcal{K'}(Q,P,t)  = \frac{1}{2} \left( P -  c(t) \right)^2 - b(t) \cos(Q).
\end{align}
\\
where $c(t) =  (\gamma/\beta) \mathbf{e}^{\beta t}$ and $b(t) = (\omega \mathbf{e}^{\beta t})^2$. Accordingly, the new canonical time is related to the original time by $d\sigma' = \mathbf{e}^{-\beta t} dt$. The Hamiltonian (\ref{K'}) describes a pendulum, whose stable fixed point shifting upwards in phase space as $c(t) \propto \mathbf{e}^{\beta t}$ and expanding as $\sqrt{b(t)} \propto \mathbf{e}^{\beta t}$. The corresponding phase-space portrait is shown in Figure (\ref{pendulum}). Both, the upwards shift and the expansion of the phase-space area, occupied by librational trajectories are necessary requirements for adiabatic capture to occur. 

The process of adiabatic capture can be understood intuitively as follows. Consider a circulating trajectory that initially resides above the pendulum's separatrix. Eventually, such a trajectory will encounter the separatrix and either enter the librational phase-space or drop down to the circulational phase-space below the separatrix. Because the phase of the encounter is essentially random, the probability of capture is simply determined by the relative rates at which the trajectory is ``invaded" by librational and circulational phase-space. In other words, $P_{\rm{capture}} = \dot{A}_{\rm{lib}}/\dot{A}_{\rm{tot}}$. The phase-space areas occupied by librating and circulating trajectories are given by $A_{\rm{lib}} = 16 \omega  \mathbf{e}^{\beta t}$ and $A_{\rm{circ}} = 2\pi (\gamma/\beta) \mathbf{e}^{\beta t} - 8 \omega  \mathbf{e}^{\beta t}$, while $A_{\rm{tot}} = 2\pi (\gamma/\beta) \mathbf{e}^{\beta t} + 8 \omega  \mathbf{e}^{\beta t}$. Consequently, 
\begin{align}
\label{Pcapt}
P_{\rm{capture}}  = 2\left(1+\frac{\pi}{4}\frac{\gamma}{\beta \omega}\right)^{-1}.
\end{align}
This expression highlights the argument made by \citet{1966AJ.....71..425G} stating that capture into resonance requires the tidal torques to be dependent on the rotation rate of the secondary (i.e. $\beta \neq 0$). Conveniently, for the tidal model of choice, $\gamma/\beta = j_2 (n-\dot{\phi})$. 

Recalling that the orbit is assumed to be circular, equation (\ref{Pcapt}) confirms that capture into the 1:1 spin-orbit resonance is certain because $j_2 = 0$. However, capture into spin-spin resonance is only certain if
\begin{align}
\label{certain}
\frac{j_2 \pi}{4 \sqrt{j_1}} \frac{\bar{r}}{R_p} \left( 1 + \frac{\dot{\phi}}{n} \right) < \sqrt{ \xi \epsilon^p \lambda_p  (\mathcal{B}_s-\mathcal{A}_s)/ \mathcal{C}_s }.
\end{align}
This leads us to conclude that capture into spin-spin resonances is likely only when the secondary's orbit is in close proximity to the primary i.e. $\bar{r}/R_p \sim \rm{few}$. However, even such a configuration requires high asphericity of both bodies. We shall examine capture into spin-spin resonances for real-life illustrative examples below. 

\section{Applications}

\begin{figure}[t]
\label{pcaptfig}
\includegraphics[width=1\columnwidth]{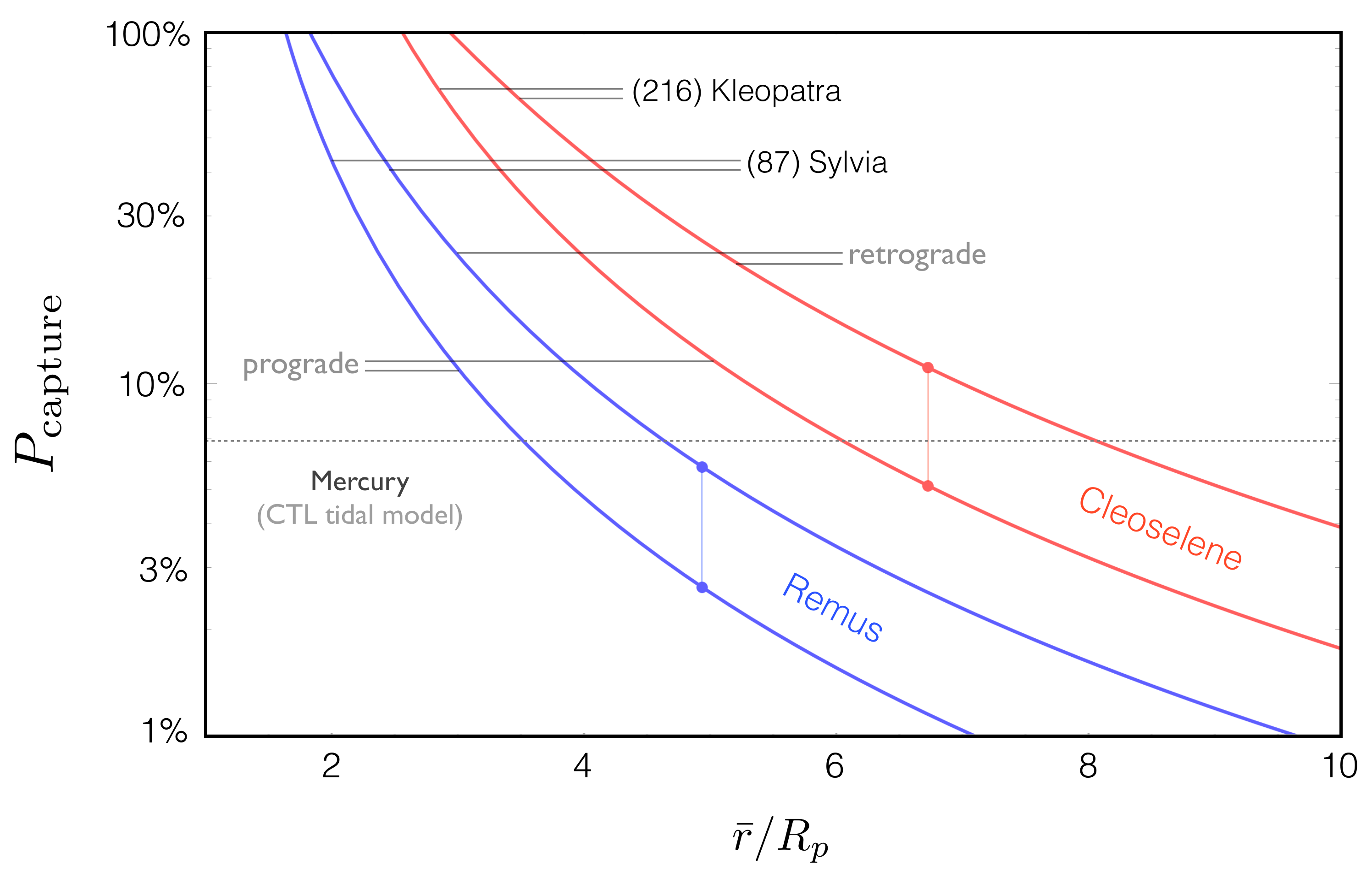}
\caption{Capture probabilities into prograde and retrograde 1:1 spin-spin resonances as a function of dimensionless radius. Best-fit parameters are used whenever available, and as before, Nix's triaxiality parameter is adopted for the satellite. The obtained probabilities are of order $\sim$ few \%, and are similar to that obtained for Mercury within the framework of the same tidal model.}
\end{figure}

Let us now examine the presented calculations within the context of observed asteroidal systems, (87) Sylvia and (216) Kleopatra. To begin with, we shall explore the rotational phase-space portrait of Remus, the inner satellite of (87) Sylvia. The physical size of Remus is $R_s \sim 7$ km, meaning that it is approximately five orders of magnitude less massive than the primary. Therefore, we can safely neglect itÕs back reaction on to the primary and take both $f$ and $\phi$ as linear functions of time. 

The $5.18$-hour rotation period of the primary is roughly $6.4$ times shorter than the orbital period of Remus. Thus, a simple way to construct a stroboscopic surface of section is to record the dynamical state $(\alpha,\dot{\alpha})$ once every 5 orbital periods or Remus, which in turn corresponds to 32 rotation periods of Sylvia. We note that another simple alternative is to section on the angle $\theta$ and plot the non-inertial phase-space $(\psi,\dot{\psi})$.

Since for the system at hand $R_s/\bar{r} \simeq 0.01$, we have chosen to only retain zeroth order (in $\epsilon$) terms in the Hamiltonian. Meanwhile, the potential of the primary is modeled as that of a triaxial ellipsoid with dimensions $384\times262\times232$ km \citep{2012AJ....144...70F}. Although the physical shape of Remus itself is observationally unconstrained, a highly irregular figure is entirely plausible given that numerous other solar system bodies (e.g. Janus, Epimetheus, Hyperion, Nix, Hydra, etc) are inferred to have $(\mathcal{B}_s-\mathcal{A}_s)/ \mathcal{C}_s \gtrsim 0.1$. For definitiveness, here we adopt the same degree of triaxiality for the secondary object as that recently derived for Nix i.e. $(\mathcal{B}_s-\mathcal{A}_s)/ \mathcal{C}_s = 0.63$ (see \citealt{ShowHam,Correia2015}).

The computed surface of section is shown in Figure (\ref{section}). The top and bottom panels depict portions of phase space occupied by prograde and retrograde spin-spin resonances respectively, whereas the middle panel shows the conventional 1:1 spin-orbit resonance. Compared with spin-orbit resonance, the widths of spin-spin resonances are diminished by a factor of 
\begin{align}
\frac{\mathcal{W}_{\rm{s-s}}}{\mathcal{W}_{\rm{s-o}}} \sim \sqrt{\frac{\mathcal{B}_p-\mathcal{A}_p}{m_p \bar{r}^2}} \leqslant \frac{R_p}{\bar{r}}.
\end{align}
For the considered case, this factor is approximately $\sim 0.1$, as can be confirmed from the Figure. An analogous surface of section can be made for the (216) Kleopatra system, however given the overall similarity in the parameters of the two systems, it would yield the same qualitative features. 

The chances that a de-spinning satellite will become permanently captured into a spin-spin resonance can be easily evaluated via expression (\ref{Pcapt}). Retaining the same degree of triaxiality as above, we have computed the probability of capture for prograde and retrograde spin-spin resonances as a function of the scaled radius for (87) Sylvia and (216) Kleopatra. The obtained curves are shown in Figure (\ref{pcaptfig}) 

As can be immediately seen, at the locations of the inner satellites, capture probabilities are generally quite low, ranging from $\sim 3\%$ to $\sim 10\%$, with the higher probabilities corresponding to the retrograde resonance (due to its somewhat enhanced width). Incidentally, these estimates are commensurate with the $\sim 7\%$ probability obtained for Mercury's capture into its current spin state, within the context of the same tidal model \citep{1966AJ.....71..425G}. 

In the case of Mercury, the effect of core-mantle friction enhances the chances of capture dramatically (albeit into the wrong resonances) \citep{PealeBoss1977,2009Icar..201....1C}. In the same vein, it is noteworthy that uncertainties persist in the understanding of tidal evolution of rubble-pile asteroids \citep{Efroimsky2015}, and it is not inconceivable that there exist physical processes that can alter our simple estimates substantially. Barring the importance of such unaccounted-for effects however, our calculations suggest that spin-spin coupling will only affect the most exotic binary/multiple objects within the sub-planetary population.

\section{Discussion}

Understanding the spin states of binary objects in the solar system is critical to disentangling their complex temporal evolution. While gravitationally bound small bodies are subject to both external and dissipative forces, the specifics of the rotational state of the binary can dictate the regime of operation of the associated effects (e.g. \citealt{GoldreichSari2009,2010Icar..207..732C}). Concomitantly, the full scope of spin dynamics of small objects has not been exhaustively explored, and remains an area of active research (\citealt{NaiduMargot,Correia2015} and the references therein). 

In this work, we have considered the rotational dynamics of highly distorted small satellites that orbit triaxial central bodies, with an eye towards identifying qualitative deviations from the conventional picture of tidal de-spinning and spin-orbit coupling. To this end, we have shown that when both bodies are sufficiently aspherical, spin-spin resonances ensue, raising the possibility that previously uncharacterized rotational behavior may be observed within the small body population of the solar system. For systems where primary spin and orbital frequencies are well separated, and satellites are negligibly small compared to their host bodies, spin-spin coupling takes on a particularly simple form, manifesting in two additional harmonics that librate when the absolute value of satellite rotation becomes commensurate with that of the primary.

For the purposes of this work, we have deliberately limited the scope of our calculations to circular orbits and only considered planar rotational motion. These simplifications allowed us to obtain a handle on spin-spin coupling from purely analytical grounds. However, in doing so, we have removed dynamical features from our description, that can exist within the framework of a more complete treatment of the problem. One such effect is YORP (e.g. \citealt{2007Icar..189..370S,2010Icar..207..732C}), associated with radiative torques exerted onto the bodies. Another effect that is well known to be consequent is chaos. 

Irregular motion in Hamiltonian systems arises from overlap of neighboring resonances \citep{1979PhR....52..263C}. Given that leading-order asynchronous spin-orbit resonance widths scale as $\propto \sqrt{e (\mathcal{B}-\mathcal{A})/\mathcal{C}}$, and high degrees of asphericity are required for spin-spin coupling to operate at a noticeable level, it is likely that for systems of interest, even small values of orbital eccentricity will immerse the vicinity of the 1:1 spin-orbit resonance into an extensive chaotic sea \citep{1984Icar...58..137W,Laskar1996}. Emblematically, this can be inferred from Figure (\ref{section}), where the separatrix of the 1:1 spin-orbit resonance extends over nominal frequencies associated with the 3:2 and the 2:1 resonances (see \citealt{MD99}). In contrast, it is possible that spin-spin resonances depicted in the same Figure will not be affected as much, since their equilibria are well removed from leading-order asynchronous spin-orbit resonances. Instead, one can speculate that accounting for finite, but nevertheless small eccentricities will simply act to engulf the separatrixes of spin-spin resonances into thin chaotic layers, in accord with the conventional modulated pendulum paradigm (see Ch. 4 of \citealt{MorbyBook}).

A related point follows regarding attitude stability. By now it is well known that satellites undergoing chaotic rotation also exhibit irregular obliquity dynamics \citep{1984Icar...58..137W,ShowHam}. On the other hand, satellites locked into stable spin resonances can retain stable spin-axis evolutions. Moreover, tidal forces typically act to damp obliquities, meaning that the end-states of quasi-periodic rotational and tidal evolution are configurations where the rotational and orbital vectors are nearly aligned. Therefore, for highly triaxial bodies, our assumption of planar rotational evolution is probably grossly violated in the vicinity of orbit-synchronous rotation, but is justified for the characterization of spin-spin resonances (which is the primary aim of this work).

Having applied conventional adiabatic capture theory \citep{1979CeMec..19....3Y,Henrard1982} to the problem at hand, we have calculated the probabilities for enduring spin-spin locking in the (87) Sylvia and (216) Kleopatra systems. While the obtained estimates are not negligibly low, they suggest that spin-spin resonant capture is generally unlikely, even for the closer satellites of these systems. Provided the limitations of the CTL tidal model employed here, these estimates deserve to be reexamined within the context of a more realistic rheology. Nevertheless, interpreting our results at face-value, we can expect that the type of rotational dynamics considered here will only impact bodies that are exceedingly closely orbiting, and will be somewhat uncommon within the overall sub-planetary census. On the other hand, unlike planets, small bodies are exceptionally numerous in the solar system, so by observationally probing ever higher magnitudes and recovering tighter binary orbits (see e.g. \citealt{PravecHarris2007}), it may be the case that the number of known bodies for which spin-spin coupling is consequential, will eventually become sizable. Accordingly, we predict that observed light-curves of such objects will reveal satellite rotation that is synchronous with the spins of the central bodies.
\\
\\
\textbf{Acknowledgments}  \\ 
We are thankful to Mike Brown for inspirational conversations and to Michael Efroimsky for providing an expedient and thorough referee report, which led to an improved manuscript.

\end{document}